# Characterization of color cross-talk of CCD detectors and its influence in multispectral quantitative phase imaging

AZEEM AHMAD[1,2,3], ANAND KUMAR[1], VISHESH DUBEY[1,2], ANKIT BUTOLA[1], BALPREET SINGH AHLUWALIA[2], DALIP SINGH MEHTA[1,4]

[1]*Department of Physics, Indian Institute of Technology Delhi, Hauz Khas, New Delhi 110016, India*
[2]*Department of Physics and Technology, UiT The Arctic University of Norway, Tromsø 9037, Norway*
[3]*ahmadazeem870@gmail.com*
[4]*mehtads@physics.iitd.ac.in*

**Abstract:** Multi-spectral quantitative phase imaging (QPI) is an emerging imaging modality for wavelength dependent studies of several biological and industrial specimens. Simultaneous multi-spectral QPI is generally performed with color CCD cameras. However, color CCD cameras are suffered from the color crosstalk issue, which needed to be explored. Here, we present a new approach for accurately measuring the color crosstalk of 2D area detectors, without needing prior information about camera specifications. Color crosstalk of two different cameras commonly used in QPI, single chip CCD (1-CCD) and three chip CCD (3-CCD), is systematically studied and compared using compact interference microscopy. The influence of color crosstalk on the fringe width and the visibility of the monochromatic constituents corresponding to three color channels of white light interferogram are studied both through simulations and experiments. It is observed that presence of color crosstalk changes the fringe width and visibility over the imaging field of view. This leads to an unwanted non-uniform background error in the multi-spectral phase imaging of the specimens. It is demonstrated that the color crosstalk of the detector is the key limiting factor for phase measurement accuracy of simultaneous multi-spectral QPI systems.

## 1. Introduction

Color charge coupled devices (CCD) have the capability to record the response of specimens as a function of red, green and blue (RGB) wavelengths in a single shot [1]. A single chip color CCD (1-CCD) camera is generally utilized to perform RGB color imaging. 1-CCD camera utilizes a pattern of Bayer color filters (25% red, 25% blue and 50% green filters) onto the image sensor to acquire RGB information of the specimen [2, 3]. Therefore, information about the object is recorded only 25% for red and blue, whereas, 50% for green wavelength. Thus, produce 50% sparse images for green wavelength, whereas, 75% sparsity is observed for red and blue wavelengths. This overall reduces the spatial resolution of the captured images [4]. The color interpolation algorithms are further employed to estimate approximate RGB value of a given pixel to produce an image having pixels equal to the number of photo sensors present in 2D area detector [5, 6]. In addition, a soft polymer dye used to fabricate Bayer color filters of the camera, which leads to the transmission of red and blue color photons into green color channel or vice versa, i.e., the response curve of these dyes are very broad [7]. The color filters and color interpolation algorithms further lead to generation of color crosstalk between RGB channels of a camera and causes noise in the color images, thus reducing the overall sensitivity of the camera [4].

Recently, white light interference microscopy (WLIM) has been implemented to extract multispectral (i.e., RGB wavelengths) phase information of the light field interrogated with the specimen simultaneously [8, 9]. Mehta *et al.* has been successfully implemented multi-spectral phase imaging of biological cells and bacteria using 1-CCD camera [9-12]. A lot of attention has been paid so far on WLIM to perform high resolution profilometry and quantitative phase imaging (QPI) of various industrial and biological samples [13-15]. However, the generation of noise level in the white light interferograms due to color crosstalk and color interpolation algorithms is not studied in depth previously. WLIM provides high spatial phase sensitivity to the phase imaging due to temporally incoherent nature of white light source [16]. Spatial phase sensitivity may be affected from the color crosstalk while using 1-CCD camera for interferometric recording due to the soft polymer dyes used in Bayer filters which have significant spectral overlap and color interpolation algorithm [3, 13]. Recently, Wang *et al* demonstrated near-field assisted WLIM for three-dimensional super-resolution microscopy [17]. Color crosstalk of CCD camera is one of the main sources of phase noise in highly sensitive multispectral QPI systems. It can also play an important role where tissue samples exhibit strong auto-fluorescence signatures at different wavelengths [18]. The main focus of this study is to provide a tool, which can accurately measure the color crosstalk of the detectors and exhibits its influence on multispectral quantitative phase imaging.

In this paper, we present a new method named interference microscopy for the characterization of color crosstalk of CCD cameras. A systematic study is done to characterize the noise level generated due to color cross talk in two different CCD cameras: 1-CCD (Infinity2-1RC) and 3-CCD (JAI AT-140GE). Mirau interferometric objective lens, which is a compact non-common path interferometer, is utilized to generate interferometric images. Three narrow bandpass filters (peak wavelength 460, 532, 620 nm having 40 nm bandwidth) are sequentially inserted into the white light beam path to record interferometric image having size M× N× P, where P =1 – 3 corresponds to red, green and blue (RGB) color channels, respectively. A set of red, green and blue (RGB) interferograms acquired from 1-CCD and 3-CCD is numerically decomposed into RGB color channels/bands. The line profiles are then plotted and compared corresponding to each bands of RGB interferograms for the measurement of color crosstalk of both the cameras. The problem of color crosstalk found to be less for 3-CCD camera as it utilizes dichroic prism optics, which has steeper spectral response, i.e., minimal spectral overlap, to split the input beam into three distinct RGB counterparts [13, 19]. In addition, 3-CCD camera does not require color interpolation algorithm to estimate the color values of a given pixel.

A theoretical framework of modified five phase shifting algorithm in the presence of color crosstalk is also presented. Further, the effect of color crosstalk on the fringe visibility of RGB channels of white light interferogram and multispectral QPI of the specimens is systematically studied both through simulations and experiments. In this work, we will discuss results, advantages, and opportunities associated with the present method.

## 2. Materials and Methods

### 2.1. Modified five frame phase shifting algorithm

Hariharan proposed a five frame phase shifting algorithm for the phase measurement of specimens [20]. The intensity modulation of five phase shifted 2D interferograms at a particular wavelength (say 620 nm) in the absence of crosstalk due to other color channels of the camera can be expressed as follows [20, 21]:

$$I_R^i(x,y) = A(x,y) + B(x,y) \cos[\varphi_R(x,y) + (i-3)\delta_R] \qquad (1)$$

where, $A(x,y)$ and $B(x,y)$ are the DC and modulation terms of the interferogram, respectively. $\varphi_R(x,y)$ is the phase information related to test object, $\delta_R$ is the phase shift between two consecutive phase shifted frames and $i$ corresponds to the number of frames ranges from 1 – 5. The phase information '$\varphi_R(x,y)$' related to test object at 620 nm can be calculated from the following expression [20, 21]:

$$\varphi_R(x,y) = \tan^{-1}\left[\sin\delta_R \frac{2(I_4(x,y) - I_2(x,y))}{I_1(x,y) - 2I_3(x,y) + I_5(x,y)}\right] \tag{2}$$

Eq. 1 can be modified into the following form if red channel interferograms are corrupted by the crosstalk due to other color channels (i.e., green and blue) of the camera:

$$\begin{aligned}I_R^i(x,y) = &\{A_R^S(x,y) + B_R^S(x,y)\cos[\varphi_R^S(x,y) + (i-3)\delta_R]\} \\ &+ \{A_G^n(x,y) + B_G^n(x,y)\cos[\varphi_G^n(x,y) + (i-3)\delta_G]\} \\ &+ \{A_B^n(x,y) + B_B^n(x,y)\cos[\varphi_B^n(x,y) + (i-3)\delta_B]\}\end{aligned} \tag{3}$$

The second and third terms in the above expression are generated due to the color crosstalk in CCD camera.

where, $A_R^S(x,y)$ and $B_R^S(x,y)$ are the DC and modulation terms of the signal interferograms corresponding to red wavelength, respectively. $\varphi_R^S(x,y)$ is the phase information related to test object at signal wavelength (620 nm). $A_G^n(x,y)$, $B_G^n(x,y)$ and $A_B^n(x,y)$, $B_B^n(x,y)$ are the DC and modulation terms of the color crosstalk noise due to green and blue color channels of the camera, respectively. $\varphi_G^n(x,y)$ and $\varphi_B^n(x,y)$ are the phase noise due to other color channels of the camera. $\delta_R$, $\delta_G$, and $\delta_B$ are the phase shift between two consecutive phase shifted frames corresponding to red (signal), green (crosstalk) and blue (crosstalk) color channels of the camera and $i$ corresponds to the number of frames ranges from 1 – 5. The phase information '$\varphi_R^S(x,y)$' related to test object at 620 nm in the presence of color crosstalk can be calculated from the following modified expression:

$$\varphi_R^S(x,y) = \tan^{-1}\left[2\sin\delta_R \frac{\{(I_4(x,y) - I_2(x,y)) - Num_{G,B}^{crosstalk}\}}{\{(I_1(x,y) - 2I_3(x,y) + I_5(x,y)) - Den_{G,B}^{crosstalk}\}}\right] \tag{4}$$

where,

$$Num_{G,B}^{crosstalk} = 2\{B_G^n(x,y)\sin(\varphi_G^n(x,y))\sin\delta_G + B_B^n(x,y)\sin(\varphi_B^n(x,y))\sin\delta_B\} \tag{5}$$

$$Den_{G,B}^{crosstalk} = 4\{B_G^n(x,y)\cos(\varphi_G^n(x,y))\sin^2\delta_G + B_B^n(x,y)\cos(\varphi_B^n(x,y))\sin^2\delta_B\} \tag{6}$$

In the absence of color crosstalk, Eq. (4) will be transformed into the expression given in Eq. (2). Similarly, for other color channels, i.e., green and blue, of the camera, the expressions of phase information '$\varphi_G^S(x,y)$' and '$\varphi_B^S(x,y)$' related to test object at 532 nm and 460 nm wavelengths, respectively, in the presence of color crosstalk due to other color channels of the camera can be written as follows:

$$\varphi_G^S(x,y) = \tan^{-1}\left[2\sin\delta_G \frac{\{(I_4(x,y) - I_2(x,y)) - Num_{B,R}^{crosstalk}\}}{\{(I_1(x,y) - 2I_3(x,y) + I_5(x,y)) - Den_{B,R}^{crosstalk}\}}\right] \tag{7}$$

$$\varphi_B^S(x,y) = \tan^{-1}\left[2\sin\delta_B \frac{\{(I_4(x,y) - I_2(x,y)) - Num_{R,G}^{crosstalk}\}}{\{(I_1(x,y) - 2I_3(x,y) + I_5(x,y)) - Den_{R,G}^{crosstalk}\}}\right] \tag{8}$$

where,

$$Num_{B,R}^{crosstalk} = 2\{B_B^n(x,y)sin(\varphi_B^n(x,y))\sin\delta_B + B_R^n(x,y)sin(\varphi_R^n(x,y))\sin\delta_R\} \quad (9)$$

$$Den_{B,R}^{crosstalk} = 4\{B_B^n(x,y)cos(\varphi_B^n(x,y))sin^2\delta_B + B_R^n(x,y)cos(\varphi_R^n(x,y))sin^2\delta_R\} \quad (10)$$

$$Num_{R,G}^{crosstalk} = 2\{B_R^n(x,y)sin(\varphi_R^n(x,y))\sin\delta_R + B_G^n(x,y)sin(\varphi_G^n(x,y))\sin\delta_G\} \quad (11)$$

$$Den_{R,G}^{crosstalk} = 4\{B_R^n(x,y)cos(\varphi_R^n(x,y))sin^2\delta_R + B_G^n(x,y)cos(\varphi_G^n(x,y))sin^2\delta_G\} \quad (12)$$

If the phase shift between two consecutive interferograms corresponds to the central wavelength in channel G is equal to $\pi/2$, (i.e., $\delta_G = \frac{\pi}{2}$), then the phase shift at peak wavelengths in channels R and B can be calculated as [13, 22]:

$$\delta_R = \frac{\lambda_R}{\lambda_G}\frac{\pi}{2}; \quad (13)$$

$$\delta_B = \frac{\lambda_B}{\lambda_G}\frac{\pi}{2}; \quad (14)$$

### 2.2. White light phase shifting interference microscopy setup

To realize the effect of color crosstalk of CCD cameras on the phase measurement experimentally, we employ WLIM depicted in Fig. 1a. The influence color crosstalk of two different cameras: 1-CCD (Infinity2-1RC) and 3-CCD (JAI GigE AT-140GE) on the fringe visibility of interferograms and subsequently on the spatial phase noise of system is studied and compared. The inset of Fig. 1 shows the source spectrum and transmission spectrum of three narrow bandpass filters having peak wavelengths 460, 532, 620 nm with ~ 40 nm spectral bandwidth.

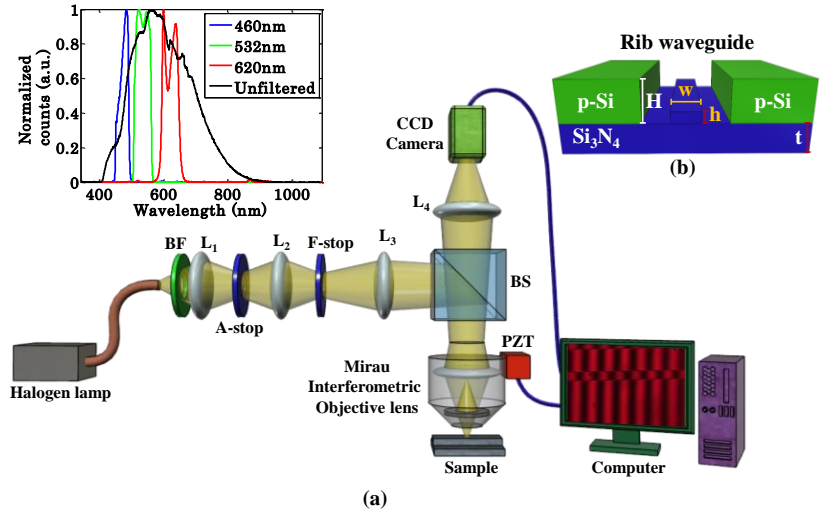

Fig. 1. (a) Schematic diagram of WL-PSIM setup. L1, L2 and L3: lenses; BS: beam splitter; A-stop: aperture stop; F-stop: field stop; BF: bandpass filter; PZT: piezo electric transducer and CCD: charge coupled device. (b) Optical rib waveguide of height 'h'.

The experimental scheme of the present setup is based on the principle of non-common path white light interference microscopy. Three narrow bandpass filters having peak wavelengths 460, 532, 620 nm with ~ 40 nm spectral bandwidth are sequentially inserted into white light beam path for the recording RGB interferograms. The image of the light source is relayed at aperture stop (A-stop) plane using lens $L_1$, where size of aperture controls spatial coherence of the light source. The source image is further relayed at the back focal plane of the objective lens with the help of lenses $L_2$, $L_3$ and beam splitter BS to achieve uniform illumination at the sample plane. The field stop (F-stop) controls the field of view (FOV) of the microscope. The beam splitter BS directs the beam towards Mirau interferometric objective lens (50×/0.55 DI, WD 3.4 Nikon, Japan) to generate white light interferograms easily. More details about the Mirau interferometric objective lens can be found elsewhere [9, 23]. The Mirau interferometer is attached with piezo electric transducer (PZT) to introduce required temporal phase shift between data frames. The phase shifted data frames are then captured using both 1-CCD and 3-CCD camera for further analysis. The modified five frame phase shifting algorithm as described in section 2.1 is utilized to understand the influence of color crosstalk on the fringe visibility of interferograms recorded by both CCD and subsequently other related studies.

### 2.3. Test object

The optical rib waveguide structure (see Fig. 1b) is used as a test specimen to quantify the color crosstalk and spatial phase sensitivity of the system. The core of rib waveguide was made of silicon nitride ($Si_3N_4$), which has a refractive index of 2.056 at 532nm wavelength. The rib waveguides were fabricated by sputtering a guiding layer of $Si_3N_4$ onto a silica (Si) substrate followed by photolithography and argon ion-beam milling [24]. The layer of $Si_3N_4$ is only partially etched down by a thickness 'h' leaving a final slab thickness of 't'. The rib waveguide had a thickness 'h' of 8nm and width 'w' of 2.5 μm. The waveguide had an absorbing layer (H = 180 – 200 nm) of poly-silicon (p-Si) onto $Si_3N_4$ as illustrated in Fig. 1b. The dimensions (height and width) of the test object have been measured using scanning electron microscopy, surface profiler and atomic force microscope. The surface profiler measurements and the atomic force microscope verified an average rib height of 8.0 nm within the noise level of the system (0.6 nm). Scanning electron microscopy measurements indicated a possible variation of ± 0.1 μm over the specified 2.5 μm wide width. More details on the optimization of waveguide fabrication can be found elsewhere [24]. The rib waveguide is placed under WLIM to record phase shifted interferograms for the measurement of color crosstalk and spatial phase noise.

### 3. Results and discussion

#### 3.1. Characterization of color crosstalk of single chip vs. three chip CCD camera

Accurate estimation of crosstalk of color CCD cameras is indispensable especially in multispectral QPI, color science and auto-fluorescence imaging of tissue samples at different wavelengths [9, 11, 18, 25]. Two different methods are presented for the characterization of color crosstalk of CCD cameras: (1) direct measurement of signal intensities at RGB color channels of bright field images and (2) interferometric measurement of signal intensities at RGB color channels of recorded interferograms corresponding to 620, 532, and 460 nm bandpass filters.

#### 3.1.1. Direct measurement

First, the color crosstalk of two different CCD cameras is measured through the direct quantification of signal intensities at RGB color channels/bands of bright field images. The bright field images are sequentially recorded by 1-CCD and 3-CCD cameras corresponding to 620, 532, and 460 nm bandpass filters. The transmission spectrum of each narrow bandpass filter is illustrated in the inset of Fig. 1. Insets of Fig. 2 depict the bright field images of a standard flat mirror corresponding to three bandpass filters. The line profiles of each RGB channels of bright field images are then plotted along the white dotted horizontal lines corresponding to 620, 532, and 460 nm bandpass filters as shown in the insets of Fig. 2. It can be clearly visualized from Fig. 2 that it is difficult to differentiate both 1-CCD and 3-CCD color cameras (compare Fig. 2a-c to Fig. d-f, respectively) on the basis of crosstalk measurement through the direct measurement of signal intensities at RGB channels of bright field images.

In the direct measurement of signal intensities at different channels of both CCD cameras, approximately constant intensity bias is observed in green and blue color channels while using 620 nm bandpass filter or vice versa (Fig. 2). The constant intensity present in green and blue channels can be easily affected from the other noises such as dark current, shot and read noise etc. of the detector [26], thus increases the measurement inaccuracy of color crosstalk.

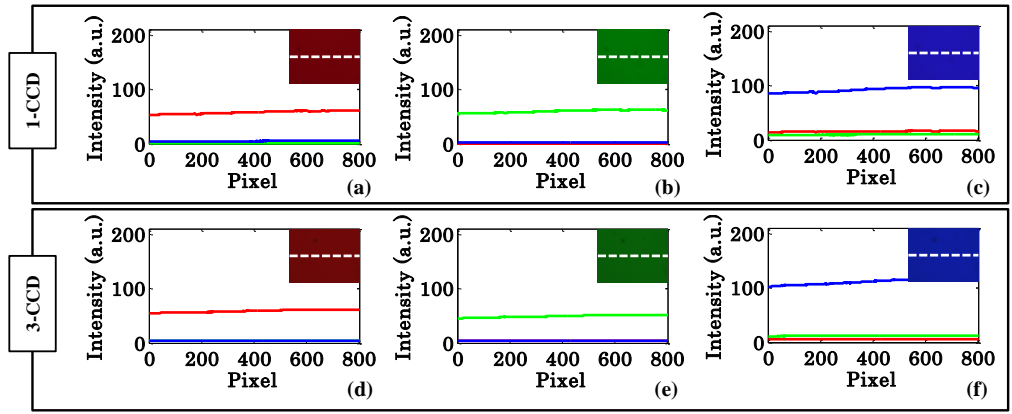

Fig. 2. Investigation of color crosstalk of 1-CCD and 3-CCD cameras obtained from the direct measurement of signal intensities at different color channels when three different bandpass color filters are sequentially inserted into the white light beam path. (a – c) Line profiles corresponding to RGB color channels/bands of the bright field images recorded by 1-CCD camera corresponding to 620, 532, and 460nm bandpass filters, respectively. (d – f) Line profiles corresponding to RGB color channels/bands of the bright field images recorded by 3-CCD camera corresponding to 620, 532, and 460nm bandpass filters, respectively. The insets of Fig. 2 are the bright field images recorded by 1-CCD and 3-CCD camera. The line profiles are plotted along white dotted lines depicted in the bright field images.

*3.1.2. Interferometric measurement*

Next, interferometric measurements of signal intensities at RGB channels/bands of the interferograms are done for the accurate quantification of crosstalk present in 1-CCD and 3-CCD cameras. For color crosstalk comparison of 1-CCD and 3-CCD camera based on interferometric measurement, three different bandpass filters are sequentially inserted into the white light beam path (Fig. 1). The RGB interferograms of standard flat mirror corresponding to each bandpass filter are then sequentially recorded by 1-CCD and 3-CCD cameras (insets of Fig. 3). Figure 3 illustrates normalized intensity line profiles of RGB color channels/bands of the interferometric images recorded by 1-CCD and 3-CCD camera corresponding to 620, 532, and 460 nm bandpass filters, respectively. The normalized intensity line profiles are plotted along white dotted lines depicted in the interferometric images (see insets of Fig. 3). For the accurate quantification of camera's color crosstalk, the minimum intensity value of each RGB color channels corresponding to a particular interferometric image (say obtained from 620 nm bandpass filter) is subtracted from its 2D intensity distribution and then normalized with respect to the maximum intensity of a color channel R. Similar steps are followed for other bandpass filters having peak wavelengths 532 nm and 460 nm for color crosstalk measurement. The flowchart for the measurement of color crosstalk is illustrated in Fig. 4. Comparison of the results depicted in Figs. 2 and 3 clearly show the advantages of interferometric measurement over direct intensity measurement for color crosstalk assessment of CCD cameras. It can be seen from Fig. 3a, the modulated intensity distribution is observed in green and blue color channels of the 1-CCD camera while using 620nm bandpass filter or vice versa, which is otherwise not present in case of 3-CCD camera (see Fig. 3d).

A constant bias (not shown in Figs. 3d – 3f) present in RGB color channels of 3-CCD camera might be due to detector's dark current noise, shot noise and read noise etc. [26]. Such color crosstalk measurement (free from other detector's noise) is not possible with the direct measurement of signal intensities at different color channels of the camera. It is worth noting that interferometric measurement of crosstalk present in color cameras is more superior and accurate over the direct measurement method presented in section 3.1.1. It is illustrated from Figs. 3a – 3c that the color crosstalk is observed between RGB color channels of 1-CCD camera, even though there is no overlap between the spectral profiles of RGB bandpass filters (inset of Fig. 1). The source of color crosstalk is due to significant RGB spectral overlap due to Bayer mosaic filter used in 1-CCD and the color interpolation algorithms, which are used to estimate approximate RGB value of a given pixel. It can be visualized from the spectral response curve of 1-CCD camera [7]. Conversely, the problem of color crosstalk is not observed while using 3-CCD cameras due to steep spectral response, i.e., slight spectral overlap, for RGB channels (see Ref. [19]]). Further, 3-CCD camera does not require color interpolation algorithm to estimate approximate RGB values of a given pixel.

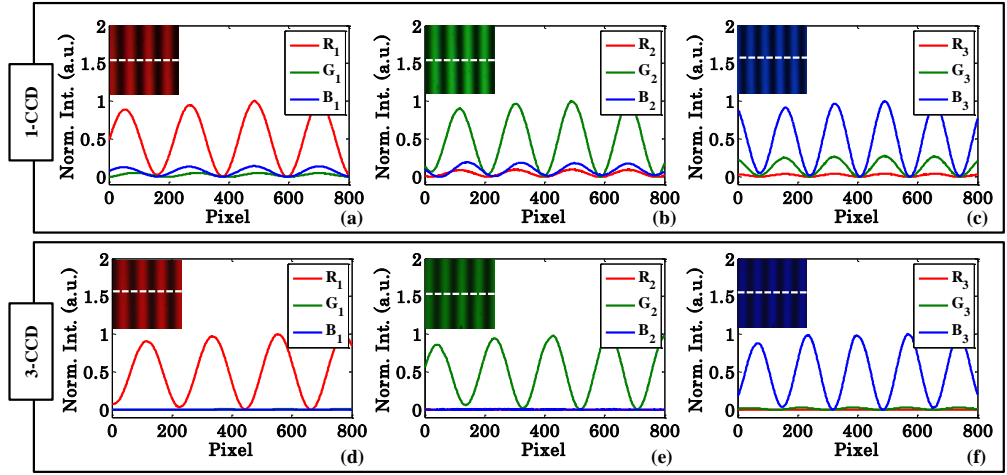

Fig. 3. Investigation of color crosstalk of 1-CCD and 3-CCD cameras obtained from the interferometric measurement of signal intensities at different color channels when three different bandpass color filters are sequentially inserted into the white light beam path. (a – c) Normalized intensity line profiles of RGB color channels/bands of the interferometric images recorded by 1-CCD camera corresponding to 620, 532, and 460nm bandpass filters, respectively. (d – f) Normalized intensity line profiles of RGB color channels/bands of the interferometric images recorded by 3-CCD camera corresponding to 620, 532, and 460nm bandpass filters, respectively. The insets of Fig. 3 are the interferometric images recorded by 1-CCD and 3-CCD camera. The normalized intensity line profiles are plotted along white dotted lines depicted in the interferometric images. The intensity of RGB color channels of each interferometric image is normalized with respect to the maximum intensity of a color channel corresponding to a particular bandpass filter.

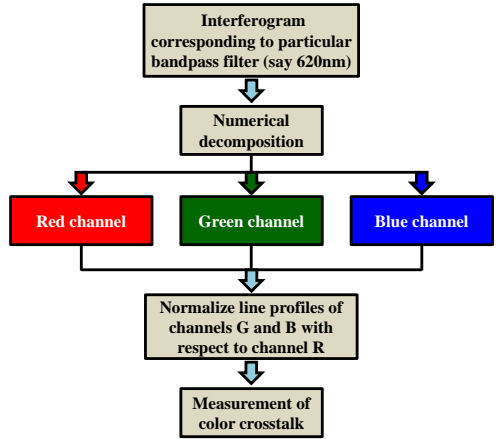

Fig. 4. Flowchart for the measurement of color crosstalk.

The normalized peak to valley (PV) intensity values in RGB color channels of 1-CCD and 3-CCD cameras are given in Table 1. These values are obtained while sequentially inserting different bandpass filters into the white light beam path. It can be envisaged from Table 2, the noise level is significantly smaller for 3-CCD camera as compared to the 1- CCD camera. The slight noise generated in RGB channels of 3-CCD camera could be due to other detector's noise [19]. It is worth noting that employment of 3-CCD instead of 1-CCD camera can be advantageous for simultaneous multi-spectral quantitative phase imaging of specimens.

**Table 1. Peak to valley intensity value generated in RGB color channels of 1-CCD and 3-CCD cameras when three different bandpass color filters having ~40nm bandwidth each at 460, 532, and 620nm central wavelengths are sequentially inserted into the white light beam path.**

| S. No. | Bandpass color filter's peak wavelength (nm) | Normalized peak to valley (PV) intensity (a.u.)$\pm 10^{-3}$ | | | | | |
|---|---|---|---|---|---|---|---|
| | | 1 – CCD color channels | | | 3 – CCD color channels | | |
| | | Red | Green | Blue | Red | Green | Blue |
| 1. | 460 | 0.044 | 0.276 | 1.000 | 0.011 | 0.048 | 1.000 |
| 2. | 532 | 0.097 | 1.000 | 0.204 | 0.004 | 1.000 | 0.004 |
| 3. | 620 | 1.000 | 0.055 | 0.143 | 1.000 | 0.008 | 0.002 |

**Table 2. Comparison of color crosstalk present in RGB channels of 1-CCD and 3-CCD camera.**

| S. No. | Color camera | Color crosstalk (%) | | | | | |
|---|---|---|---|---|---|---|---|
| | | Channel R | | Channel G | | Channel B | |
| | | Due to Channel G | Due to Channel B | Due to Channel B | Due to Channel R | Due to Channel R | Due to Channel G |
| 1. | 1-CCD | 9.7 % | 4.4 % | 27.6 % | 5.5 % | 14.3 % | 20.4 % |
| 2. | 3-CCD | 0.4 % | 1.1 % | 4.8 % | 0.8 % | 0.2 % | 0.4 % |

The normalized PV intensity values presented in Table 1 are further utilized for the quantification of signal to noise ratio (SNR) of individual color channels using the following expression:

$$SNR_{channel\ R,G,B}(dB) = 20\ log\left(\frac{signal_{channel\ R,G,B}}{noise_{crosstalk}}\right) \quad (15)$$

For the measurement of $SNR_{channel\ R}$ using Eq. 15, the normalized PV intensity value of channel R is considered $signal_{channel\ R}$, whereas, noise is considered the PV intensity value of the leakage in channel R, which is obtained from the interferograms corresponding to the bandpass filters having 532 and 460 nm peak wavelengths. Similarly, measurement of $SNR_{channel\ G}$ and $SNR_{channel\ B}$ of other color channels G and B is also performed. Table 3 presents SNR of different color channels of 1-CCD and 3-CCD camera under different color crosstalk noise levels. It can be noticed from Table 3 that the SNR is high for 3-CCD camera as compared to the 1-CCD camera. For uncertainty analysis about the crosstalk measurement, experiments were repeated ten times under similar experimental conditions. The uncertainty in SNR measurement is found to be less than 1dB.

Table 3. Signal to Noise ratio (S/N) of RGB channels of 1-CCD and 3-CCD camera under different color crosstalk noise levels.

| Signal to Noise ratio (S/N) | | 1-CCD (dB) | 3-CCD (dB) |
|---|---|---|---|
| **Channel R** | $SNR = 20\, log\left(\frac{S_{channel\,R}}{N_{channel\,G}}\right)$ | 20.21 | 46.92 |
| | $SNR = 20\, log\left(\frac{S_{channel\,R}}{N_{channel\,B}}\right)$ | 27.03 | 39.04 |
| | $SNR = 20\, log\left(\frac{S_{channel\,R}}{N_{channel\,G} + N_{channel\,B}}\right)$ | 16.95 | 36.09 |
| **Channel G** | $SNR = 20\, log\left(\frac{S_{channel\,G}}{N_{channel\,B}}\right)$ | 11.17 | 26.27 |
| | $SNR = 20\, log\left(\frac{S_{channel\,G}}{N_{channel\,R}}\right)$ | 25.06 | 41.31 |
| | $SNR = 20\, log\left(\frac{S_{channel\,G}}{N_{channel\,B} + N_{channel\,R}}\right)$ | 9.57 | 24.86 |
| **Channel B** | $SNR = 20\, log\left(\frac{S_{channel\,B}}{N_{channel\,R}}\right)$ | 16.87 | 51.56 |
| | $SNR = 20\, log\left(\frac{S_{channel\,B}}{N_{channel\,G}}\right)$ | 13.79 | 46.93 |
| | $SNR = 20\, log\left(\frac{S_{channel\,B}}{N_{channel\,R} + N_{channel\,G}}\right)$ | 9.18 | 42.92 |

*3.2. Influence of color cross talk on the fringe visibility and multispectral QPI*

To understand the influence of color crosstalk on the fringe visibility of white light interferogram's RGB components, we performed both the simulation and the experimental studies. Subsequently, its effect on the multispectral quantitative phase measurement is studied and found to be crucial for 1-CCD color camera.

*3.2.1. Simulation study*

*3.2.1.1. Effect on fringe visibility and fringe width*

Simulation study is performed to understand the effect of color crosstalk on fringe visibility and fringe width of the interferogram. For all simulation studies related to crosstalk, we considered 460 nm, 532 nm and 620 nm peak wavelengths with zero spectral bandwidth for the generation of 2D monochromatic interferograms. However, spectral bandwidth cannot be equal to zero for the experimentally recorded RGB interferograms into different color channels of the camera. Bandwidth depends on the spectral response curve of the CCD camera [19]. In simulation, bandwidth is considered to be equal zero for clearly understand the influence of crosstalk on fringe visibility and fringe width of the interferogram. 2D interferograms of the simulated donut shape phase object are generated with and without the presence of color crosstalk. Figures 5a and 5d illustrate the 2D interferograms with and without the presence of color crosstalk, respectively. These simulated interferograms correspond to 460 nm wavelength, i.e., the blue color channel (say) of color CCD camera. Figure 5d is generated by considering the presence of 10% crosstalk noise due to the red and blue color channels both. The line profiles along blue dotted lines are depicted in Figs. 5b and 5e corresponding to the interferograms shown in Figs. 5a and 5d.

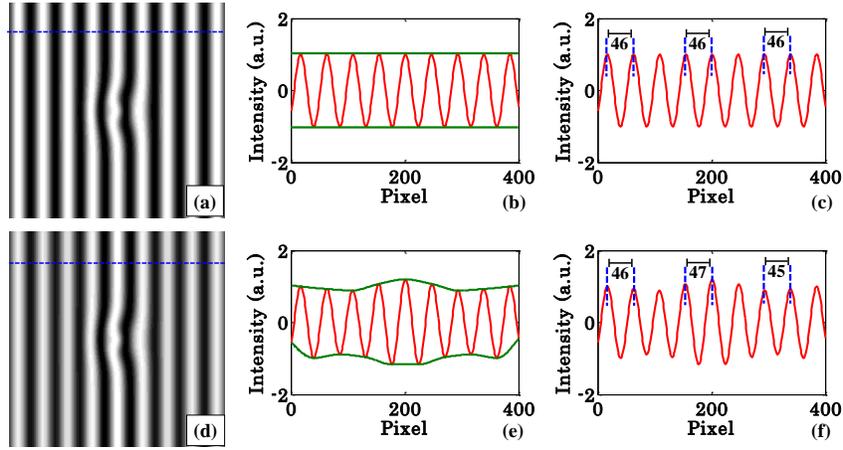

Fig. 5. Influence of color crosstalk on the fringe visibility and fringe width of interferogram. (a, d) Interferograms corresponding to 460 nm wavelength without and with crosstalk. (b, e) Corresponding line profiles along blue dotted horizontal lines with visibility profile shown in solid green color lines. (c, f) Representations of equal and spatially varying fringe width of interferograms in the absence and presence of crosstalk, respectively.

To clearly envisage the effect of color crosstalk on fringe visibility, an algorithm is developed using MATLAB to draw the envelope of interference fringes. The green color solid lines depicted in Figs. 5b and 5e represent the variation of fringe visibility over the width of interferograms. The fringe visibility is found to be constant for the crosstalk free interferogram, whereas, it varies irregularly over the interferogram's width in the presence of 10% crosstalk (Figs. 5b and 5e).

Next, the influence of crosstalk on the fringe width of the interferogram is studied. It can be seen in Fig. 5c that if crosstalk is zero then a constant fringe width equal to 46 pixels is observed everywhere over the width of interferogram. However, the presence of 10% crosstalk due to red and green channel into blue color channel (say) causes unwanted fringe width change across the width of interferogram (Fig. 5f). The unequal fringe width varying from 45 pixels to 47 pixels is observed over the interferogram's width. This leads to the generation of non-uniform background in the reconstructed phase image corresponding to 460 nm wavelength. Similar results are also obtained through simulation studies for other color channels such as red and green channel. Since, the spatial frequency of crosstalk generated due to red and green channel is different from the spatial frequency of blue channel. Therefore, addition of crosstalk noises (red and green channels) into signal (blue channel) lead to the unequal fringe width of the corresponding interferogram as illustrated in Fig. 5f.

### 3.2.1.2. Effect on multispectral quantitative phase imaging

After systematic investigation of the effect of color crosstalk on the fringe visibility and fringe width of interferograms, the simulation studies are performed to understand its effect on the multispectral quantitative phase imaging. To study the color crosstalk, five $\delta_G = \pi/2$ phase shifted interferograms corresponding to green color channel are simulated and then 10% crosstalk noise due to red (phase shift $\delta_R = \frac{\lambda_R}{\lambda_G}\frac{\pi}{2}$) and blue (phase shift $\delta_B = \frac{\lambda_B}{\lambda_G}\frac{\pi}{2}$) color channels is added. Similarly, for red and blue color channels, five equal phase shifted interferograms having phase shift $\delta_R = \frac{\lambda_R}{\lambda_G}\frac{\pi}{2}$ and $\delta_B = \frac{\lambda_B}{\lambda_G}\frac{\pi}{2}$, respectively, with crosstalk effect are simulated. Different phase shifts (i.e., $\delta_R$, $\delta_G$, $\delta_B$) between the consecutive phase shifted frames for red, green and blue color channels is introduced to imitate the experimental conditions.

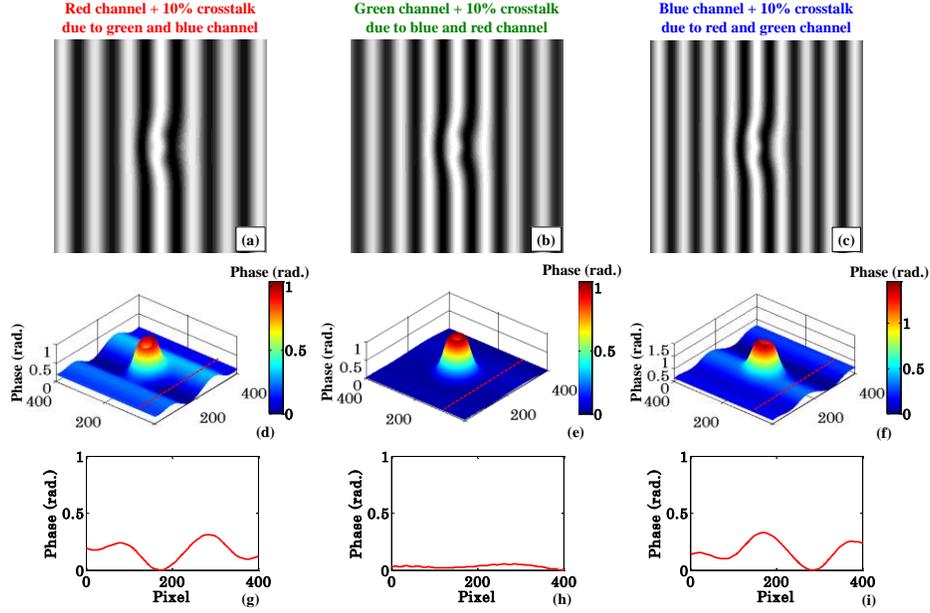

Fig. 6. Influence of color crosstalk on Multispectral QPI. (a – c) One of the five phase shifted interferogram corresponding to red, green and blue color channels in the presence of 10% crosstalk due to the respective other two channels. (d – f) Recovered phase maps of simulated phase object for red, green and blue color channels, respectively. (g – i) Line profiles plotted along the red dotted line shown in Fig. 6d – 6f. The colorbar is in rad.

One of the five phase shifted interferogram corresponding to red, green and blue color channels in the presence of 10% crosstalk due to the respective other two channels are exhibited in Figs. 6a – 6c, respectively. Five frame phase shifting algorithm given in Eq. 2 is utilized for the phase recovery of simulated phase object as illustrated in Fig. 6d – 6f. It is worth noting that the presence of crosstalk introduces a non-uniform background in the reconstructed phase maps and add unwanted measurement error to the phase values. The slight difference in the behavior of non-uniform background obtained from simulation and experiments could be due to zero bandwidth consideration of light source during simulation. The line profiles along red dotted lines shown in the recovered phase images are then plotted and depicted in Fig. 6g – 6i. The line profiles clearly demonstrate the generation of non-uniform background due to color crosstalk in the recovered phase images.

### 3.2.2. Experimental investigation

### 3.2.2.1. Effect on fringe visibility

To visualize the influence of color crosstalk on fringe visibility experimentally, white light interferograms are sequentially recorded using 1-CCD and 3-CCD camera as shown in Figs. 7a and 7c, respectively. Bandpass filters are not inserted into the beam path for the recording of white light interferograms. Each color interferogram is then decomposed into its monochromatic constituents, i.e., RGB components, using MATLAB. Figures $7a_1$-$a_3$ and $7c_1$-$c_3$ represent RGB components of color interferograms recorded by 1-CCD and 3-CCD camera, respectively. The line profiles corresponding to RGB components for both CCD cameras are further plotted as illustrated in Figs. $7b_1$-$b_3$ and $7d_1$-$d_3$. To clearly envisage the effect of color crosstalk on fringe visibility, an algorithm is developed using MATLAB to draw the envelope of interference fringes. The black color solid lines depicted in Figs. $7b_1$-$b_3$ and $7d_1$-$d_3$ represent the variation of fringe visibility for RGB components over the sensitive area of CCD cameras. The fringe visibility of the RGB components over the 1-CCD sensor's active area changes in an irregular manner (see Fig. 7b$_1$-b$_3$). This is found to be in a good agreement with the simulation results presented in Figs. 5d-5f. However, for 3-CCD camera the fringe visibility does not vary significantly and approaches towards the ideal one as shown in Fig. 5b.

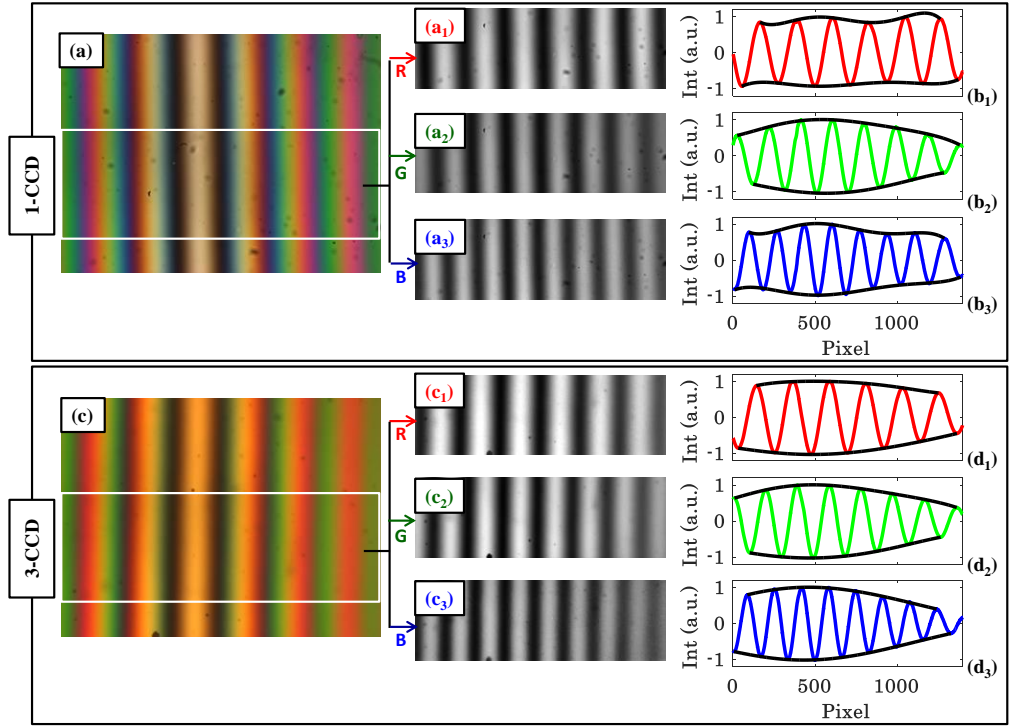

Fig. 7. Effect of color crosstalk of 1-CCD and 3-CCD camera on the fringe visibility. The narrow bandpass color filters are not used into the white light beam path for the recording of color interferograms. Black color lines represent the visibility curve profiles o09f the monochromatic (RGB) constitutions of white light interferograms for both 1- CCD and 3-CCD.

*3.2.2.2. Effect on multispectral quantitative phase imaging*

It has been clearly demonstrated that the color crosstalk greatly influence the fringe visibility profile of the monochromatic constituents (RGB components) corresponding to the white light interferogram acquired from 1-CCD camera compared to 3-CCD camera. In addition, color crosstalk of CCD camera affects the multispectral QPI of the specimen significantly. To understand the effect of crosstalk in QPI, experiments are conducted on rib waveguide and corresponding five equal phase shifted white light interferograms are recorded by sequentially employing 1-CCD and 3-CCD camera. The phase shift between interferograms is introduced with the help of PZT. Each white light interferogram is decomposed into its monochromatic constituents, i.e., RGB components, for multispectral phase recovery by employing Eq. 2. The modified phase recovery algorithm (Eq. 4) is not implemented because the measurement of crosstalk terms $Num_{G,B}^{crosstalk}$ and $Den_{G,B}^{crosstalk}$ given in Eqs. 5 and 6 is not straightforward.

The recovered phase images of rib waveguide corresponding to RGB channels of 1-CCD and 3-CCD camera are illustrated in Figs. 8a-8c and Figs. 8g-8i, respectively. The backgrounds of the recovered phase maps are non-uniform and add unwanted phase measurement error while using 1-CCD camera for interferometric recording. The non-uniform background is in accordance with the simulation results as shown in Fig. 6. It can be observed that the noise is more prominent in case of blue channel.

The experiment conducted with 3-CCD camera does not have color crosstalk issue between RGB color channels. Therefore, the non-uniform background is not observed in case of 3-CCD camera as depicted in Figs. 8a-8c. However, the background modulation of very small amplitude having twice spatial frequency to the original spatial frequency of the monochromatic interferograms corresponding to RGB color channels of 3-CCD camera is observed. This could be due to the minute unequal phase shift error between the consecutive phase shifted interferogram as shown in the previous studies [20].

The line profiles along red and green dotted lines (Figs. 8a-8c and Figs. 8g-8i) for 1-CCD and 3-CCD camera are illustrated in Figs. 8d-8f and Figs. 8j-8l, respectively. It can be seen from the line profiles (red dotted lines in Fig. 8) that higher non-uniformity in the background is observed using 1-CCD than that obtained with 3-CCD camera. Further, the line profiles along green dotted lines for 1-CCD and 3-CCD camera are plotted for the measurement of multispectral phase maps of rib waveguide as presented in green color profiles in Figs. 8d-8f and Figs. 8j-8l, respectively. The phase value of the waveguide's absorbing layer (180 – 200 nm, H in Fig.1) corresponding to RGB wavelength using 1-CCD and 3-CCD camera are given in Table 4.

**Table 4. Multispectral phase measurements of rib waveguide while employing 1-CCD and 3-CCD camera for white light interferometric recording.**

| Camera | Phase (rad.) | | |
|---|---|---|---|
| | **Red channel** | **Green channel** | **Blue channel** |
| **1-CCD** | 1.934 $\pm$ 0.018 | 2.248 $\pm$ 0.016 | 2.535 $\pm$ 0.032 |
| **3-CCD** | 2.005 $\pm$ 0.013 | 2.293 $\pm$ 0.014 | 2.530 $\pm$ 0.037 |

The presence of non-uniform background in phase images (say corresponding to 620 nm) of rib waveguide (Fig. 1b) obtained from 1-CCD camera is due to the leakage of green and blue color photons into red channel of 1-CCD camera (see Fig. 3). Spatial frequency of the interferogram for a fixed angle between object and reference beam varies as a function of wavelength due to the change in OPD. For higher wavelength (620 nm), low spatial frequency of the interferogram over the FOV of camera is obtained, whereas, high spatial frequency is obtained for lower wavelength (460 nm) as depicted in Fig. 3. It is worth noting that the leakage of red color photons into green and blue color channels produces modulated crosstalk noise with same spatial frequency as that for signal, i.e., red color channel. Similarly, the spatial frequency of modulated crosstalk in blue and red channel due to green color photons is found to be same as signal in green channel. It is observed from simulation that the presence of only 10% crosstalk noise in red color channel (say) due to green and blue color channels slightly changes the fringe width of the interferogram across the FOV of camera. This leads to non-uniform background in the reconstructed phase maps of the specimens. Similar results are also observed for green and blue color channels of 1-CCD camera.

However, 3-CCD camera free from color crosstalk, does not produce such non-uniform background in the reconstructed multispectral phase maps. The large spatial phase noise is observed in case of blue color channels of both 1-CCD and 3-CCD camera. This could be due to the photon noise as halogen lamp spectrum (inset of Fig. 1) has lesser number of blue color photons into the spectral bandwidth range of soft polymer dye (for 1-CCD) and dichroic coating (for 3-CCD) compared to other color channel's spectral response (peak wavelengths ~ 532nm and 620nm). The spectral response of both cameras can be found in Refs. [7, 19].

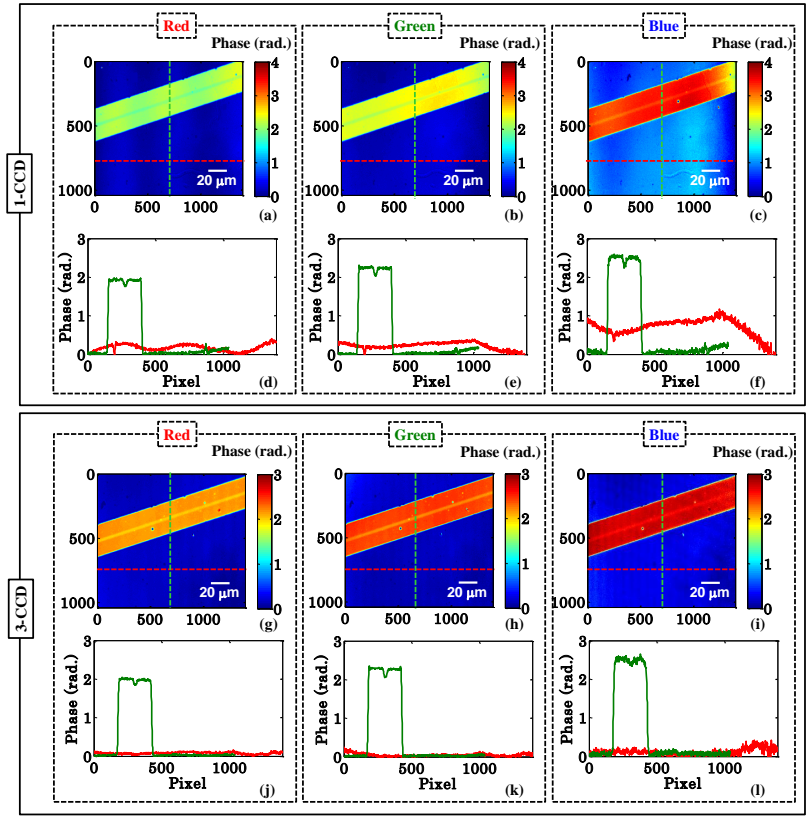

Fig. 8. Multispectral quantitative phase imaging of rib waveguide recovered from white light interferogram recorded with 1-CCD and 3-CCD camera, respectively. (a-c) Recovered pseudo 3D phase map of rib waveguide obtained from RGB color channels 1-CCD camera, (d-f) corresponding line profiles along red dotted horizontal lines illustrating non-uniform background and green dotted vertical lines depicting the inverse phase profiles of rib waveguide. (g-i) Recovered pseudo 3D phase map of rib waveguide obtained from RGB color channels 3-CCD camera, (j-l) corresponding line profiles along red dotted horizontal lines illustrating absence of non-uniform background and green dotted vertical lines depicting the inverse phase profiles of rib waveguide. The color bars represent phase in rad.

## 4. Conclusion

Here, we proposed a compact white light interference microscopy setup for the accurate measurement of color crosstalk generated in RGB color channels of a CCD camera. The color crosstalk generated in RGB color channels while using 1-CCD and 3-CCD camera for multispectral imaging, is systematically studied and compared. Narrow bandpass filters at central wavelengths 460, 532, and 620 nm with ~ 40 nm bandwidth each, are sequentially inserted into the white light beam path to measure color crosstalk of both cameras. The influence of color crosstalk on the fringe visibility of RGB color channels is then studied for both CCD cameras. The bandpass filters are not inserted into the white light beam path to investigate the effect of color crosstalk on the variation of fringe visibility and fringe width over the sensor's active area. The visibility curve for 3-CCD camera is found to be close to the ideal one. The variation in the fringe visibility profiles is arises due to the low temporal coherence length of the filtered white light beam at the dichroic prism optics.

Further, the effect of color crosstalk is investigated on the multispectral quantitative phase imaging. The five frame phase shifting algorithm in the presence of color crosstalk of CCD camera is modified and its mathematical formulation is developed. However, the modified five frame phase shifting algorithm is not implemented for the crosstalk free multispectral phase recovery of specimens because the accurate estimation of crosstalk terms is not straightforward, e.g., estimation of $Num_{G,B}^{crosstalk}$ and $Den_{G,B}^{crosstalk}$ terms in the red color channel of the camera or vice versa. This can be explored in future.

It is observed that the presence of color crosstalk introduces non-uniform background in the reconstructed multispectral phase images and subsequently phase measurement accuracy of the system while using 1-CCD camera. This is also confirmed from the simulation studies. The influence of color crosstalk from recovered multispectral phase images is minimized by employing 3-CCD camera for interferometric recordings. As it does not suffer from the problem of color crosstalk, therefore, noise terms $Num_{G,B}^{crosstalk}$ and $Den_{G,B}^{crosstalk}$ are found to be nearly equal to zero for 3-CCD camera. Thus, the modified five frame phase shifting algorithm given in Eq. 4 is transformed into Eq. 2 for 3-CCD camera and leads to crosstalk free phase recovery.

The developed algorithm with some advancement can be further employed for the multi-spectral quantitative phase imaging of various industrial and biological specimens. The benchmarking of color crosstalk in the detector will be useful for multi-spectral quantitative phase imaging of tissues sections that exhibit strong auto-fluorescence signatures at different wavelengths [27]. The present study may also find useful applications in pulse oximetry, currently, being used to provide important information about tissue perfusion, oxygenation and potentially function during surgery [28, 29].

**Funding:** The authors are thankful to Department of Atomic Energy (DAE), Board of Research in Nuclear Sciences (BRNS) for financial grant no. 34/14/07/BRNS. B.S.A acknowledges the funding from the European Research Council, (project number 336716) and Norwegian Centre for International Cooperation in Education, SIU-Norway (Project number INCP- 2014/10024). D.S.M acknowledges University Grant Commission (UGC) India for joint funding.

**Acknowledgement:** Authors would like to acknowledge Jean-Claude Tinguely and Firehun T. Dullo for their help with the waveguide chip. This work has made use of the Spanish ICTS Network MICRONANOFABS partially supported by MEINCOM.